\documentclass[letterpaper]{article}

\usepackage{natbib,alifeconf}  
\usepackage{amsmath}
\usepackage{url,hyperref,cleveref}
\usepackage{booktabs}
\usepackage{xcolor}

\newtheorem{prop}{Proposition}

\newtheorem{definition}{Definition}

\title{Does Capital Dream of Artificial Labour?}
\author{
    Marcin Korecki$^{1,*}$ \and 
    Cesare Carissimo$^{2,*}$, 
    \mbox{}\\
    $^1$TU Delft, the Netherlands \\
    $^2$ETH Zurich, Switzerland \\
    $^*$Equal Contribution \\
    mkorecki@tudelft.nl
} 

\begin{document}
\maketitle

\begin{abstract}
This paper investigates the concept of Labour as an expression of `timenergy' — a fusion of time and energy — and its entanglement within the system of Capital. We define Labour as the commodified, quantifiable expansion of timenergy, in contrast to Capital, which is capable of accumulation and abstraction. We explore Labour's historical evolution, its coercive and alienating nature, and its transformation through automation and artificial intelligence. Using a game-theoretic, agent-based simulation, we model interactions between Capital and Labour in production processes governed by Cobb-Douglas functions. Our results show that despite theoretical symmetry, learning agents disproportionately gravitate toward capital-intensive processes, revealing Capital’s superior organizational influence due to its accumulative capacity. We argue that Capital functions as an artificially alive system animated by the living Labour it consumes, and question whether life can sustain itself without the infrastructures of Capital in a future of increasing automation. This study offers both a critique of and a framework for understanding Labour’s subjugation within the Capital system.

\end{abstract}

Spectres of automation hang in the air driven by the logics of Capital and the energy of Labour. There is a clear tendency of technology to automate many facets of human life. Fundamental sectors of the economy like agriculture and manufacturing, are transformed by autonomous machines which shift the requirements of human labour. Also transformed, are essential relations between people mediated by algorithms. It seems that many skills which set us apart from inert matter can now be encoded in artificial labourers. We are in an unprecedented position to wonder whether a system of Capital which organises Labour needs humans at all.

Is Capital pushing inevitably towards the automation of all Labour? Essential to this query is to ask what it is that labour needs to be Labour, and whether Labour is fundamentally related to life. It seems, for example, that as an inescapable part of nature, life toils ceaselessly. In as much as this comes in the form of expansion of energy in time it might already be apparent at the level of a single cell, which the observer distinguishes from inert matter by the activity they interpret as purposeful. This activity remains present at subsequent levels of life's organization. The living organs formed by cells perform specialized work that maintains the vital functions of an organism. The bodies formed by organs move around in a total commitment to the thermodynamic principle. Ants build anthills, beavers dams, bees hives but it is the humankind's Sisyphean efforts of which we can speak with most certainty. And what are humanity's efforts primarily if not the survival, the homeostatic prerogative that ensures the body is nourished and sheltered\footnote{`Labour is, first of all, a process between man and nature, a process by which man, through his own actions, mediates, regulates, and controls the metabolism between himself and nature.' \citep{marx2019capital}}. Cast as means of survival, human labour appears above all a necessity -- a necessity that humanity has, for centuries, struggled to diminish through automation. 

Our presently achieved automation (2025 AD) appears to profoundly enhance our ability to survive, but does not free people from the necessity to work. Humans still labour even though we increasingly have artificially capable machines that turn energy into useful work. This mystery can be illuminated by the realization that just as we attempt to bring forth machine servants we have already enslaved ourselves to the mechanisms required to produce the machine servants. Thus, the biologically alive human forms a hybrid complex \citep{baltieri2023hybrid} with the artificially alive Capital \citep{carissimo2024capital}. We argue that it is Capital and its accumulation, that drives humanity into their labours\footnote{When referring to Capital in this work we follow the definition from \citep{carissimo2024capital} according to which Capital is a historical agential system that arises and continues embedded in the complexity of a social sphere. As a system it is embodied in quantified value and that which pursues its maximization.
}. Labour is a crucial activity in the Capital system. The abstract nature of Capital finds its realisation, its embodiment in Labour and ultimately it is this artificially alive Capital that humanity is subservient to. 

We believe that the hybrid Capital-Labour system ought to be studied as a form of artificial life. In this paper we make a contribution by building on previous characterizations of Capital as Artificial Intelligence \citep{carissimo2024capital}, proposing new definitions, and constructing a learning agent based model of Capital and Labour allocation to study their structural differences. 

\subsection{Background}

The theories of Labour are perhaps as varied as the theories of Capital. No consensus has been reached as to the drivers of Labour or its valuation. The attitude to Labour in ancient times appears less than enthusiastic\footnote{`The Greeks in their era of greatness had only contempt for work: their slaves alone were permitted to labour: the free man knew only exercises for the body and mind... The philosophers of antiquity taught contempt for work, that degradation of the free man, the poets sang of idleness, that gift from the Gods.'  \citep{lafargue2012right}}. The existence of explicitly coercive Labour in the form of slavery and the lowly status of slaves paints a clear picture of the ancient attitudes to Labour. As such, even technically voluntary (though perhaps motivated by poor socio-economic conditions) Roman wage labourers were seen as not much different than slaves \citep{cicero1870officiis}. 

More positive theoretical attitudes towards Labour started emerging much later. Locke, for example would come to associate Labour with private property \citep{locke2016second}, which then evolved into the Labour Theory of Value of \citet{adam2016wealth} and \citet{marx2019capital}. Much like Capital, Labour has also been classified and categorized into a variety of types, such as abstract and concrete. The common ground for the early theories of Labour appears the association of Labour and Capital in the process of production. 

The modern character of the forces of Capital brought forth by mercantilism, industrial revolution and eventually capitalism has been associated with a particular work ethics of the populations that were at the vanguard of this acceleration. This protestant work ethic, emphasising dedication to one's labour and its value in itself, has been claimed to be at the heart of modern-day capitalism \citep{weber2013protestant}. Along with the breakdown of the feudal stratification of society, these historic processes revolutionized the social division of Labour into one that would become progressively market driven. As a result the modern attitude to work is a mixed one, with significant criticism present, some authors calling for abolitions of work \citep{black1985abolition, fischer2022let}, other praising idleness \citep{russell1972praise}, and still others praising the market's efficiency at organising Labour \citep{hayek1975kinds}.  

The unholy marriage of Labour and Capital is a frequent theme in Marxist or Anarchist literature. From these perspectives Labour, instrumentalised in the service of Capital, is ultimately detrimental to the subjects that engage in it. The alienating nature of Labour, its mental and physical precariousness and the inherent propensity for exploitativeness have all been a common theme of the critique of the contemporary capitalism. Some authors argue that voluntary labour is an oxymoron as it always occurs under a form of coercion \citep{lordon2014willing}. The hierarchical, asymmetrical relationships between humans (boss and employee, capitalist and labourer) induced by the organisation of labour have also come under increasing criticism \citep{graeber2007manners}.


\subsection{This Paper}

In the present work we will follow a pragmatic definition of Labour that embeds it in the Capital system. But first, we must discuss the resource available to us humans which is truly limited above all, and this is `timenergy', the fusion of time and energy which we borrow and build upon \citep{timenergy}.

\begin{definition}
    Timenergy is reliable, reusable and routinely available large blocks of energy-infused time. It is the fundamental resource, combined time and energy that can then be used for any intentional activity \citep{timenergy}.
\end{definition}

\citet{timenergy} argues that timenergy is the most fundamental and limited resource available to us as people. He also argues that `the good life' is a life with timenergy. We never get it back, but it replenishes. As we will see, timenergy is what constrains the productivity of Capital. When we labour for Capital, our timenergy is reduced to Labour.

\begin{definition}
Labour is an agential activity that takes the form of a quantified expenditure of energy in time. As such it is a commodified form of timenergy that is embedded in the Capital system. Any purposeful activity taking place within the Capital system is Labour. 
\end{definition}

This subject has a choice as to how their timenergy will be spent \footnote{We note the philosophical debate on agency, intentionality, and directedness, which we wish to stray clear of, but refer the interested reader to a useful characterization of agency for the economic and artificially intelligent context \citep{martinelli2023toward}.}. Why then does it commit its timenergy to Capital? Especially, since this commitment is one of de-personalization, as all timenergy is equal in the eyes of Capital\footnote{`Capital no longer recruits people, but buys packets of time, separated from their interchangeable and occasional bearers. De-personalised time has become the real agent of the process of valorisation, and de-personalised time has no rights, nor any demands either. It can only be either available or unavailable, but the alternative is purely theoretical because the physical body despite not being a legally recognised person still has to buy his food and pay his rent.' \citep{berardi2005info}}. Why does life follow that which is only artificially alive\footnote{`Far from being an internal property or quality of labour, productivity indexes the dehumanization of cyborg labour-power. As regenerative commoditization deploys technics to substitute for human activity accounted as wage costs, it obsolesces the animal, the organism and every kind of somatic unity, not just in theory, but in reality; by tricking, outflanking and breaking down corporeal defences. The cyborg presupposes immunosuppression.' \citep{land1995meat}}? 

We consider a system where agents possessing timenergy have already decided that it is in their best interest to engage with a system of Capital. First, we propose a series of propositions describing the role of Labour in the Capital system. We identify an asymmetry between Labour and Capital in that the former cannot be accumulated while the latter can. Following that, we analyse the incentives of Capital and Labour in production. Subsequently, we design multi-agent learning simulation experiments that shine further light on the matter. All in all, we ask what is the fundamental difference between Capital and Labour and how does the asymmetry between them influence production dynamics?

\section{Labour in Capital}

In this section we posit the following propositions constituting a general model of Labour as embedded in the Capital system. 

\begin{prop}
\label{prop:discrete}
    Labour is fundamentally discretized (digital) through the units of capital. 
\end{prop}

Labour is fully quantified in so far as it is a part of the Capital system. As the medium of Capital is quantified value it is itself perpetuating a universal medium of commensurability in which Labour becomes readily discretized into digital units of capital. 

Furthermore, Labour seen in terms of the timenergy that has been committed to the Capital system can also be quantified (e.g. two hours of energy-infused time). Whenever we talk about this timenergy cost of Labour we refer to it as labour units to distinguish it from the capital units that can be used to indicate the value given to Labour in the Capital system.

\begin{prop}
\label{prop:timenergy}
    Labour is upper bounded by the agent's timenergy. 
\end{prop}

A given agent's timenergy is a finite resource. The amount of energy an agent can expand in a given period of time is bounded. Thus, Labour as a commodified form of timenergy is likewise bounded. This limitation corresponds to the thermodynamic constraints of a biological agent or mechanical constraints of an artificial agent. 

\begin{prop}
\label{prop:finite}
    Labour cannot be accumulated. 
\end{prop}

Labour, being bounded cannot be accumulated. As Labour is a form of quantified timenergy its limitations are that of time and energy. Time is always running out and so clearly cannot be stored, while energy can be saved but cannot be increased beyond a certain limit. In a sense energy-in-time is an inflationary resource that when preserved is still undergoing a constant decay.  

\begin{prop}
\label{prop:partition}
    Labour is directed at the accumulation of capital units. 
\end{prop} 

Labour, in so far as it lends itself to quantification, is universalized by it as a pursuit of capital units. As such any labour must as its aim have the maximization of capital units available to the agent. 

\begin{prop}
\label{prop:action}
    Labour belongs to the set of actions that may be afforded by the units of capital. 
\end{prop}

Capital units afford actions (and observations) to the agents. Labour is one type of action that can be afforded to the agents by units of capital. Thus, as an action, Labour presupposes the existence of units of capital.  

\begin{prop}
\label{prop:generation}
    Labour actions can generate new units of capital, combined with units of capital in production processes.
\end{prop}

Labour can generate new units of capital for instance through a production process, which takes a given amount of units of labour and a given amount of capital units and returns some amount of new units of capital. In the rest of the paper, we focus on production processes as functions of the form $P(C,L) = C'$, where $C$ are capital units and $L$ are labour units, though in general a production process may take more and varied inputs, and produce more and varied outputs.

Labour and Capital come together in production processes to generate capital units. We take the production to be the key area in which the interaction between Labour and Capital emerge. In the rest of this paper we address the following research question: how does the asymmetry between Capital and Labour, that Capital can be accumulated but Labour is limited and fixed, influence production dynamics?

\section{Analysis of Production Incentives}

In this section we fix the functional form of production to analyse the resulting incentives for agents that control both capital units and labour units. A standard way of representing production takes the form of a Cobb-Douglas production function with constant return to scale: a function 
\begin{equation}\label{eq:cobbdoug}
    P(C,L) = MC^\beta L^{1-\beta},
\end{equation}
where $C$ is total capital input, $L$ is total labour input, $M$ is an output multiplier, and $\beta$ is the elasticity of capital -- the dependence of the process on capital units relative to labour units.

Given a production function of the form in \autoref{eq:cobbdoug}, how would agents in control of both capital and labour units allocate their resources? Assuming that an agent has a choice between two processes, each with their own elasticity parameter $\alpha,\beta \in [0,1]$ and multiplier $M_i\in [1,\inf)$ for $i\in[k]$, we can analyse the first derivatives of the production function to calculate the marginal productivities of capital ($mpC$) and labour ($mpL$):

\begin{align*}
    mpC &= \frac{\delta Y}{\delta C} = (1-\beta)M (\frac{C}{L})^{\beta}, \\
    mpL &= \frac{\delta Y}{\delta L} = (\beta)M (\frac{L}{C})^{1 - \beta}.
\end{align*}

These marginal productivities, as they are called in economics, can also be thought of as a cost of capital and a wage rate, for capital units and labour units respectively. They can also be thought of as the instantaneous capital reward and labour reward. The multiplier $M$ has the most interpretable effect -- both marginal productivities are linear and monotonically increasing in $M$, so processes with higher multipliers are preferable. In the rest of this section, and the rest of this paper, we assume that the multipliers are the same for all processes.

The marginal productivities are influenced by the capital elasticity $\alpha, \beta$ and the ratios of labour and capital units $\frac{L}{C}$. Agents that seek to maximize their capital return from production will select the resource and process that yields the highest return on the resource. If we assume an agent which controls an infinitesimal resource, we can conclude that the agent would pick the resource and process with the highest marginal productivity. By analysing the marginal productivities we can see that:
\begin{enumerate}
    \item If a capital unit moves from a process A to a process B, 
    \begin{itemize}
        \item the capital reward increases at A and decreases at B.
        \item the labour reward decreases at A and increases at B.
    \end{itemize}
    \item If a labour unit moves from a process A to a process B, 
    \begin{itemize}
        \item the capital reward decreases at A and increases at B.
        \item the labour reward increases at A and decreases at B.
    \end{itemize}
\end{enumerate}

Important in these relationships is the asymmetry where capital rewards increase relatively in processes with increasing labour ratios $\frac{L}{L+C}$, and oppositely, labour rewards increase relatively in processes with increasing capital ratios $\frac{C}{L+C}$. 

\begin{figure}
    \centering
    \includegraphics[width=\linewidth]{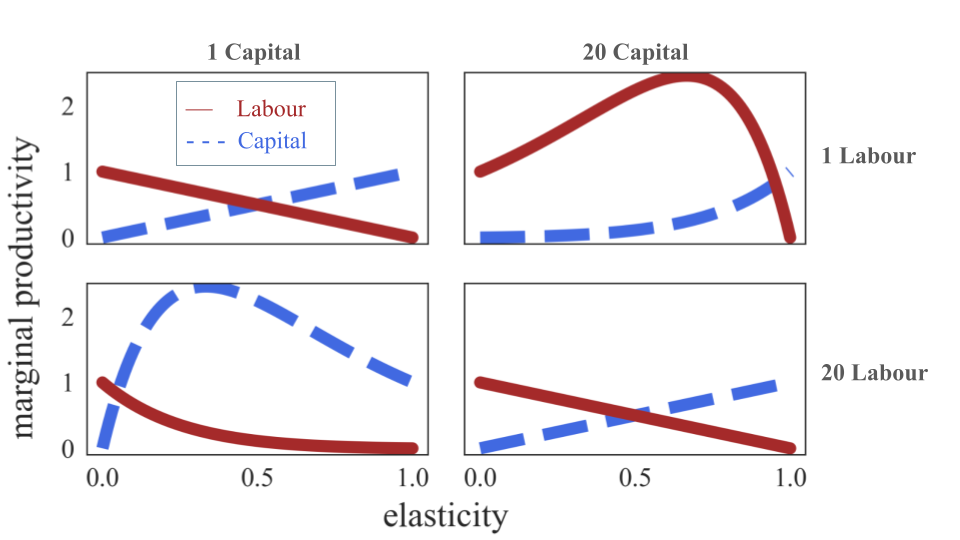}
    \caption{The Marginal Productivities of capital and labour as a function of the elasticity of capital for different ratios of capital-to-labour units allocated to a given process.}
    \label{fig:marginal-productivity}
\end{figure}

Capital chases labour units, and Labour chases capital units. However, Capital and Labour tend to prefer opposite values of elasticity. This is seen in \autoref{fig:marginal-productivity}, where the capital and labour rewards are plotted for varying labour ratios, and varying elasticity $\beta$. We can see that, fixing a capital-to-labour ratio e.g. $1:1$, the maximum value of marginal productivity does not align for Capital and Labour. In other words, Capital and Labour tend towards different processes. 

Moreover, for the extreme case of $1:20$ Labour to Capital, we can see a counter-intuitive result of lower elasticities giving higher marginal productivity for Capital. When there is much more labour units, it makes sense for the capital agent to assign its capital units to processes whose return depends more significantly on the number of labour units assigned. The high elasticity processes would give a smaller return as they depend on labour units (which are in this case abundant) to a lesser extent. 

From this analysis we have emphasized three main features of the incentive structure of the production functions: 
\begin{enumerate}
    \item Labour is attracted to processes which have relatively more capital units than labour units allocated.
    \item Capital is attracted to processes which have relatively less capital units than labour units allocated.
    \item The ideal process to which labour agents would allocate their resources and the ideal processes to which capital agents would allocate their resources have different elasticities.
\end{enumerate}

This appears to pose a coordination problem to the agents engaged in the Game as some number of labourers and capitalists need to select the same process for it to produce returns. However, they are incentivised to prefer different elasticities. 

\section{Computational Setup}

In this section we put together a capital-labour production game as an agent based model played by learning agents. Our approach follows similar attempts of using agent based models \citep{birner2024breaking}, simulations \citep{kant2020worksim} and especially learning agents \citep{korecki2023artificial} to study social and economic systems. The idea of applying alife methods to the study of economies and the emergence of certain economic relations has a long history \citep{ tesfatsion2002agent, tesfatsion2018economists} in the field and aligns well with our goal.

\subsection{The Game}

To further study the production process in an experimental setting we create a simple instantiation of Capital and Labour as a stochastic game. This game is played by $n$ agents, with $k$ production processes in $P$. Each production process has a distinct elasticity parameter $\beta_p$ for $p\in P$. Each agent has an initial amount of capital units $c$, and timenergy $\tau$. Their capital units can vary over time based on their actions. On the other hand, timenergy is not accumulated, but is replenished every turn. Each turn, agents decide whether to allocate their capital or to allocate their timenergy to a production process. Agents receive capital units as a reward for their allocations, which depends on the production process and the allocations of all other agents.

The output $Y=P(C,L)$ is then redistributed to the agents following a redistribution rule. The capital providers split $\beta Y$ proportional to their capital allocation, and the labourers split $(1-\beta)Y$ proportional to their timenergy allocation \footnote{We assume that every unit produced is bought by the market at a price of 1.}.

\subsection{The Agents}

The agents in our model interact without directly modelling each other. Therefore, they create a non-stationary environment for each other. We implement the agents as stateless reinforcement learners that learn the values of process-resource combinations. We pick $Q$-learning as the reinforcement learning method of the agents \citep{watkins1992q}, as it is well suited to non-stationary rewards, and follow inspiration from the continual learning setup of \cite{carissimo2024counter}. Each agent has a $Q$-table that stores the $q$-values for every process-resource combination (agents can commit either its labour or capital units to a given process). At each turn they pick an action by referring to their $Q$-table following an $\epsilon-$greedy action policy. The exploration rate parameter $\epsilon \in [0,1]$ is the probability with which agents explore -- pick a process-resource combination uniformly at random from the available process-resource combinations. Otherwise, with probability $(1-\epsilon)$ they pick the process-resource with the maximum $q$-value.

After all agents pick their process-resources, each agent receives a (possibly different) reward $r_i$ given their process-resource $a_i$. The agents then update their $q$-value for the action $a_i$ following the equation: $Q(a) = Q(a) + \alpha (r + \gamma max_{a'}Q(a') - Q(a))$. The learning rate parameter $\alpha$ adjusts the rate at which new reward information $r$ changes the $q$-values. The discount factor $\gamma$ adjusts the extent to which agents care about future rewards.

\subsection{Experiments}

\begin{center}
\begin{table}[]
\begin{tabular}{cc|cc}
 Game & Values & Learning & Values \\ 
  \hline
  \hline    
  $\beta$ (elasticity) & $[0.1 - 0.9]$ & $\alpha$ & $[0.01, 1]$\\
  $M$ (multiplier) & $1$ &  $\gamma$ & $[0, 0.99]$ \\
  $k$ (processes) & $2^n, n \in [1, 8]$ & $\epsilon$ & $[0, 0.1]$ \\
  $n$ (agents) & $2^n, n \in [2, 10]$ & \\
 \end{tabular}
 \caption{The Game and Learning parameters and their values as used in the experiments.}
 \label{tab:params}
\end{table}
\end{center}
\begin{figure}
    \centering
    \includegraphics[width=\linewidth]{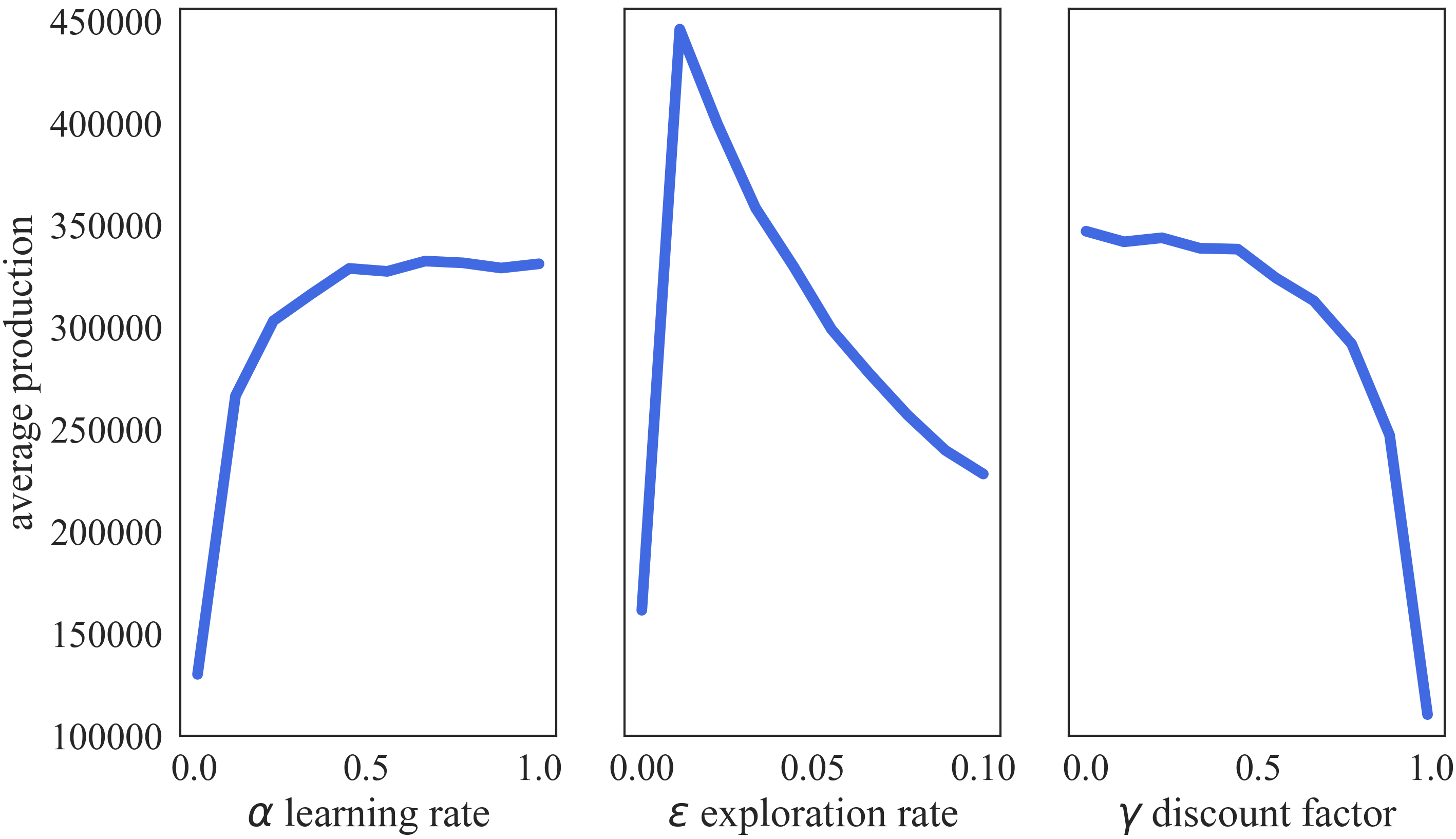}
    \caption{The influence of individual learning parameters on aggregate average production. The plotted levels of production average over all other experimental parameters as specified in \autoref{tab:params} to emphasize the parameter's more general effects.}
    \label{fig:learning}
\end{figure}
We run a series of simulation experiments using the game and agents as described above\footnote{Code available at: \href{https://github.com/CCarissimo/capital}{https://github.com/CCarissimo/capital}}. Agents all start with equal endowments of capital units, $100$, and equal amounts of labour units, also $100$. Their $q$-values are initialized uniformly at random in the range $[0,100]$ for each process. Each turn, they decide whether to allocate all of their capital units or all of their labour units to one of $k$ production processes. At the end of the turn, agents get rewarded in capital units. Labour units reset at the start of each turn to $100$. Capital units can be accumulated turn after turn.

We measure the following aggregate values from all of our experiments, averaging over all parameter combinations:
\begin{enumerate}
    \item average production: an average over all production processes and over all learning steps.
    \item capital strength: the elasticities of processes are drawn uniformly at random from the range $[0.1,0.9]$. For each simulation run, we check which process produced the most, and compare its elasticity $\beta_*$ to the maximum available elasticity $\beta_{max}$. The capital strength is then the normalized distance from the maximum elasticity $\frac{\beta_* - \beta_{min}}{\beta_{max}-\beta_{min}}$.
    \item labour ratio: is the average number of labourers over time divided by the total number of agents.
\end{enumerate}

The parameters we use for our experiments are reported in \autoref{tab:params}. The learning parameters are 10 equidistant values in the range specified in \autoref{tab:params}. For the case where the number of processes equals $1$, we select 9 equidistant process elasticities from the range specified in \autoref{tab:params}. For cases where the number of processes is greater than $1$, we sample $9$ arrays of process elasticities uniformly at random from the same range. For each combination of parameters specified in \autoref{tab:params} we run a simulation for $5000$ learning steps, and repeat this simulation $4$ times. Finally, we add an extensive ablation of learning parameters to ensure the robustness of our results.

The specific effects of learning parameters on average production can be seen in \autoref{fig:learning}, where we average over all other parameter settings. Our results show that higher $\alpha$ and lower $\gamma$ lead to higher average production. The $\epsilon$ at 0.02 yields the highest production. In the rest of our results we average all values over all of the learning parameters to provide the most robust estimates of the relationships we are interested in without bias towards specific regions of the learning parameter space.

\section{Results}

\begin{figure}
    \centering
    \includegraphics[width=\linewidth]{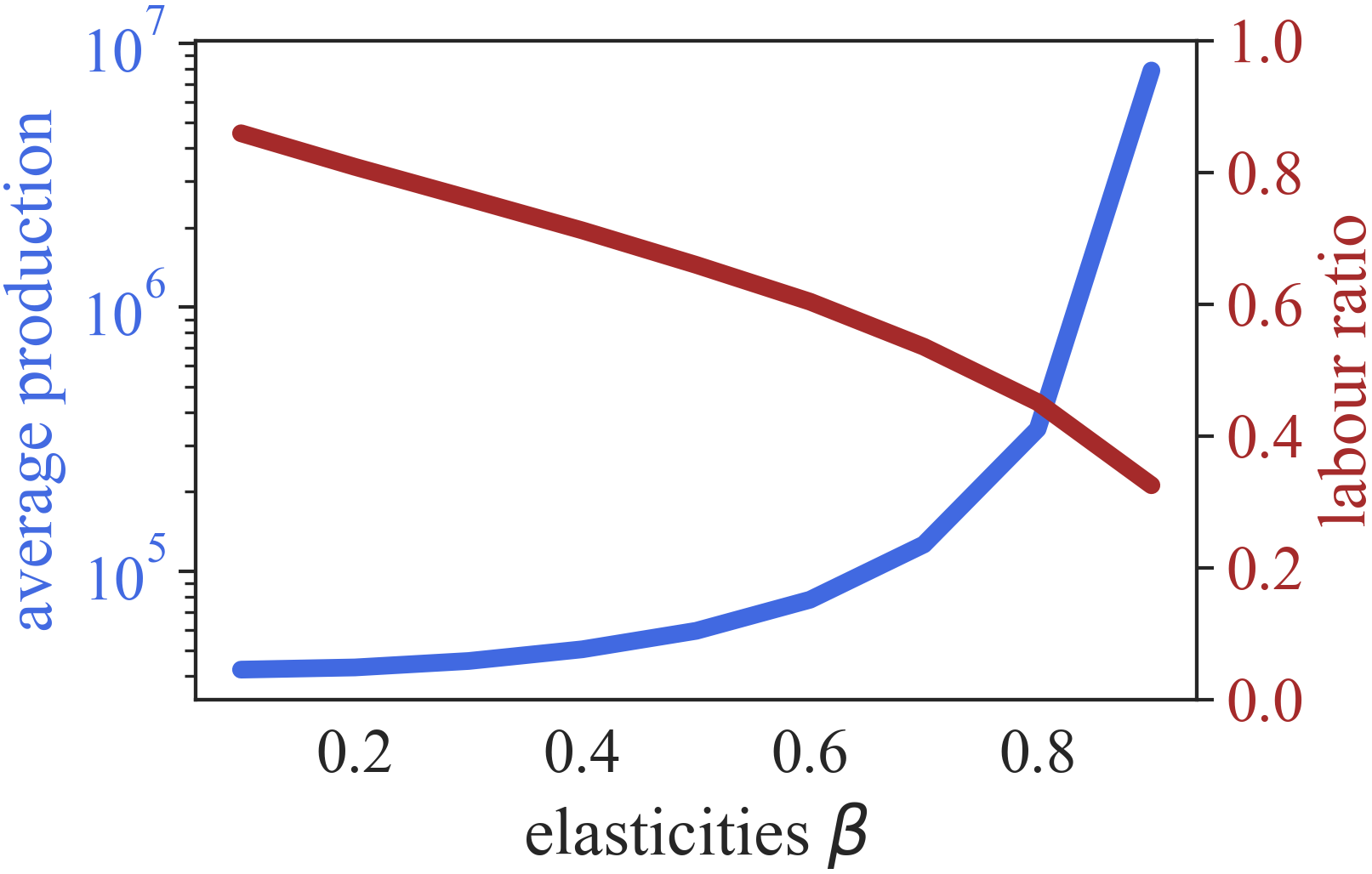}
    \caption{The average production and labour ratio as functions of the process elasticity.}
    \label{fig:production}
\end{figure}

\autoref{fig:production} gives a sense of the relationship between the elasticity value of a given process and the average production of that process. The higher elasticities lead to higher average production. We also note that the labour ratio (labourers to all agents) decreases as the elasticity increases. This can be interpreted as an effect of a growing automation of production. Processes with high elasticity need fewer labourers to achieve the same production (given high levels of capital) and so can be seen as more automated, requiring less living labour, than processes with low elasticities. 

\begin{figure}
    \centering
    \includegraphics[width=\linewidth]{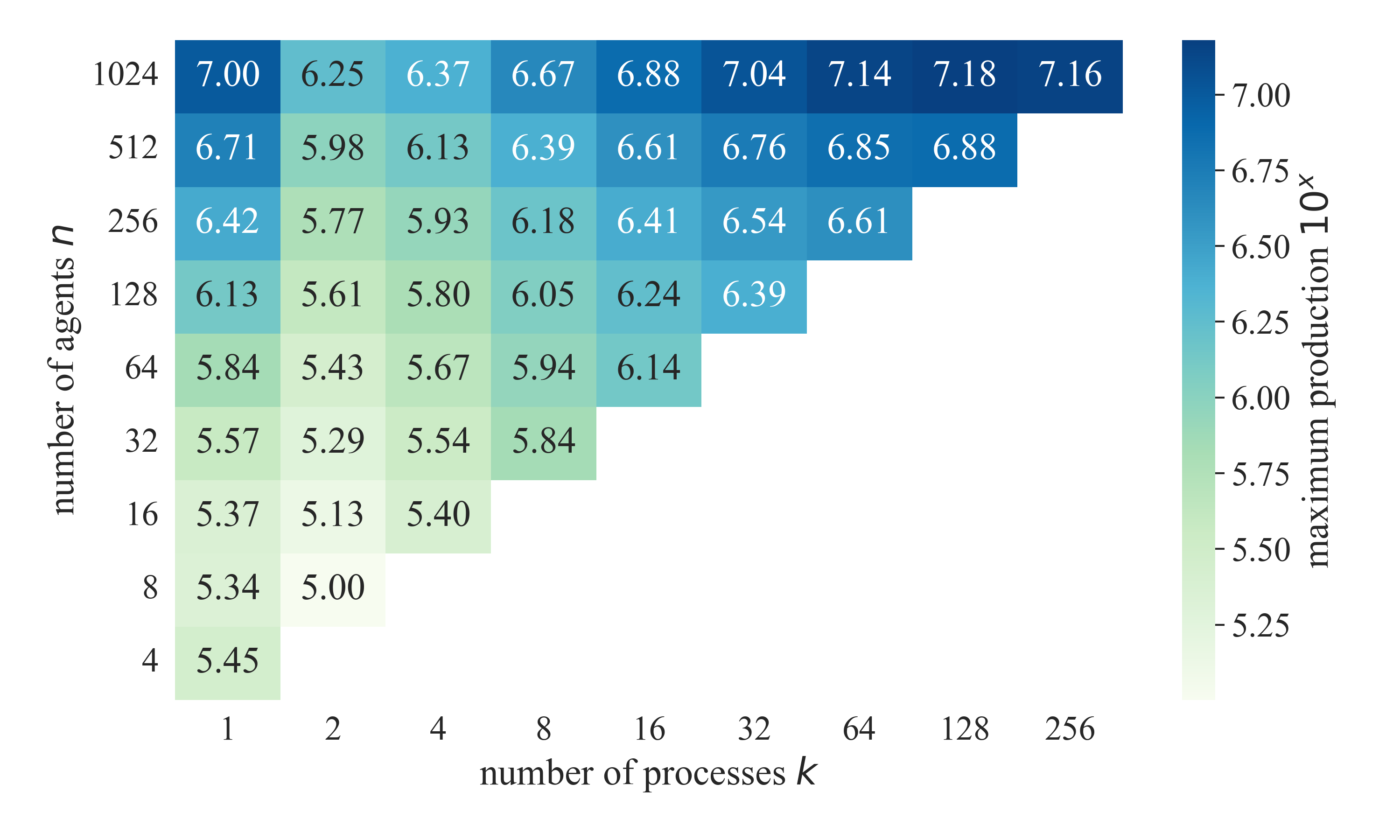}
    \includegraphics[width=\linewidth]{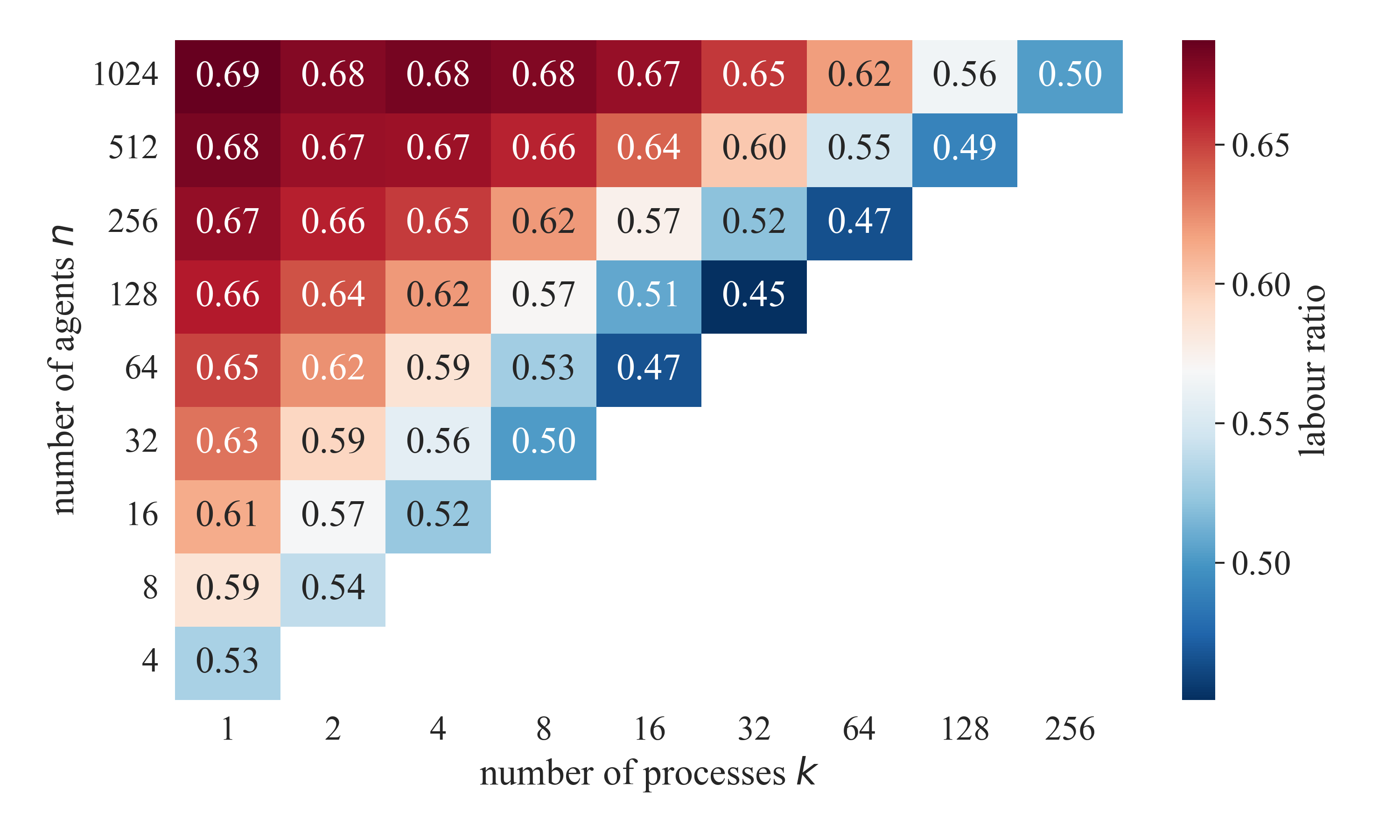}
    \includegraphics[width=\linewidth]{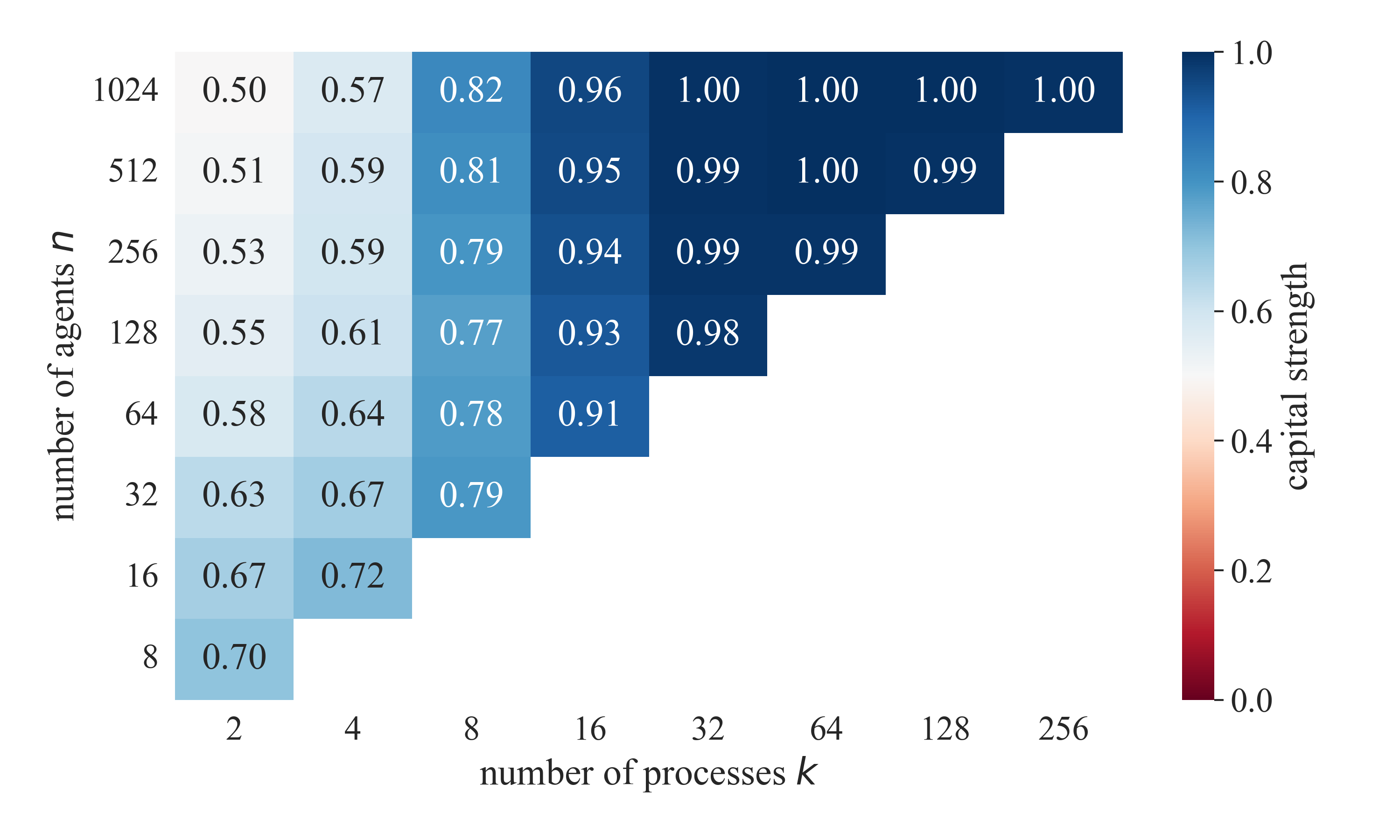}
    \caption{Heatmaps of metrics as a function of the number of processes (horizontal axis) and the number of agents (vertical axis). The maximum production to the number of agents and processes (top). Values of maximum production are on a log scale. The Labour Ratio is the ratio of the average number of labourers to the average number of capitalists (middle). The capital strength is the distance of the elasticity of the process that produced most, from the process with the highest elasticity, normalized between the maximum and minimum elasticities available (bottom). The results for $k=1$ are removed as the notion of capital strength is not applicable to this case.}
    \label{fig:heatmaps}
\end{figure}


In \autoref{fig:heatmaps} (top) we report the maximum value of production for different numbers of agents and processes. We can see that the maximum production increases with the number of processes as well as with the number of agents. The explanation for an increase of production in the number of agents is clear, since the total number of agents increases the total number of available labour units which are the `bottleneck' of production. On the other hand, as the number of processes increases there is a higher chance that processes with higher elasticity will be drawn, and as we see in \autoref{fig:production}, higher elasticities lead to higher average production, and lower labour ratios. Indeed this is confirmed also in \autoref{fig:heatmaps} (middle), the heatmap for labour ratios -- the labour ratio decreases in the number of processes, as the likelihood of a high elasticity process increases. Interestingly, we find that as the number of agents increases, the labour ratio also increases, while keeping the number of processes fixed. 


The final confirming result of our analysis comes from the capital strength, reported in \autoref{fig:heatmaps}  (bottom). We see that, indeed, as the number of processes increases, the elasticity of the most productive process increases all the way to the maximum. As the number of processes increases, the difficulty of coordination increases too, as it is less likely that just by chance capital units and labour units will be allocated to the same process. This interacts with the previous observation that the probability of encountering a high elasticity process also increases with the number of processes.



In \autoref{fig:heatmaps} (middle) we can see that the ratio of labourers to all agents is strongly dependent on the number of agents (the more agents the higher the ratio). Moreover, it is also dependent on the number of processes -- the more processes available the lower the labour ratio.


\section{Conclusion}

In this paper, we have defined Labour as living in Capital. Our model assumes that production depends on inputs of capital units and labour units. The key difference between capital units and labour units stands in their accumulation; while capital units can be accumulated and generated with ease, labour units cannot be accumulated. This breaks the symmetry between Capital and Labour and enables Capital to direct, organize, and exploit Labour. In fact, capital units stand as a store of labour power -- `dead labour'\footnote{`There is a familiar humanist response to this becoming zombie at the limit possibility of the modern worker, which is associated above all with the word alienation. The processes of de-skilling, or ever accelerated re-skilling, the substitution of craft by abstract labour, and the increasing interexchangability of human activity with technological processes, all accompanied by the dissolution of identity, loss of attachment, and narcotization of affective life, are condemned on the basis of a moral critique. A reawakening of the political is envisaged, aimed at the restoration of a lost human integrity. Modern existence is understood as profoundly deadened by the real submission of humane values to an impersonal productivity, which is itself comprehended as the expression of dead or petrified labour exerting a vampiric power over the living.'
\citep{land1988kant}} -- which enables an agent with capital units to decide where other agents allocate their labour units. Thus, the answer to the main question we posed is that Labour follows Capital because of it is possible to accumulate Capital, while it is not possible to accumulate Labour. Timenergy needs to be used or otherwise will be lost but to use one's timenergy to perform Labour allows it to be stored in Capital by proxy. 

As such we have substantiated the value of the model proposed in \citet{carissimo2024capital} and used the concept of the living Labour to ground the abstract model of the Capital system in a Game that lends itself to experimentation and further study.

Furthermore, in \citet{carissimo2024capital} it has been argued that the quantification Capital produces is not inherently meaningful. However, Labour could be seen as this aspect of the system that indeed carries meaning with it as long as it is provided by living beings. The living Labour is what carries the potential to infuse the quantification performed by Capital with meaning. It is precisely these kinds of hybrid systems which may present us a most compelling epistemology of Artificial Life \citep{pattee1995artificial}, where Capital provides an artificial structural logic and living beings, nature itself, provides the timenergy required.

What we also hope to have emphasized is the real relevance of Capital as artificial life, where it is embedded in a social sphere with Labour that clearly lives. It is the Labour that animates Capital making it artificially alive, more, than would be for just an abstract system of measurement, which Capital is. Capital is the software, Labour is the hardware. But the biologically alive labour, human or animal, that drives the Capital system forward is not its necessary component. As we have seen over the recent decades, the accelerating automation of Labour is threatening to make it independent of any living intercession. This is the apocalyptic vision of accelerationism of Nick Land, where a fully automated, autonomous, artificial Labour force replaces the biological Labour. Capital might reach a point where it does not need life to supply it with Labour but will life survive without Capital? 

\subsection{Limitations}

Our work is not without its limitations, while allowing us to give tentative answer to our research question it does not claim to be complete. For one we do not consider the effects of demand which would affect the value of production. This demand, driven by desires of agents partaking in the Capital system, is an important consideration in asking the question whether or not Capital can escape humans or whether the value that is created by Capital depends on humans. After all, Capital creates the means of survival for many humans and it can be hypothesized that should humans be replaced, Capital would still need to provide some means of survival or preservation for whatever process would provide it with its energy.

Moreover, in our model we assume that the agents have already decided to commit their timenergy to the Capital system via Labour. However, it is also clear that timenergy can be expanded in pursuits, which technically can occur beyond the Capital system (even if the sphere free from Capital might be progressively diminishing due to its expansive nature). Activities crucial to human flourishing such as leisure, companionship, dreaming or even non-remunerated work are all possible without the intercession of Capital but still timenergy is required to engage in them. Since we have focused on the study of the Capital system, these activities are missing from our model, but that should in no way be taken as a statement as to their value and importance. 

Finally, we are missing an in depth uncertainty analysis. We run each parameter combination 4 times and report the averages, and note that further statistical test would need more repetitions to yield statistical significance. Nonetheless, the standard deviations of our means, are heteroscedastic, large for the maximum production values, and small for the labour ratio and capital strength.

\subsection{Future Work}

How can we really address the question of why humans continue to work while Capital continues to produce more advanced automation? How can the model we have be extended to include the desires of agents and individuals in a meaningful way to study the relationship between productivity, value and desire? These questions edge closer and closer to the problem of agency, which is a common theme in artificial life research \citep{moreno2005agency, rohde2009agency, froese2023irruption}.

The meaning that life brings, and its potentially intimate relationship with value in the Capital system presents another avenue open to further investigations. What happens to the meaning in the hypothetical, fully automated, breakaway Capital system, where there is no more life left to assign meaning to its products and quantifications? Realizing the challenging and complex nature of these questions we do not attempt to give definite answers to them in this work but remain committed to their further study in the future.  

\section{Acknowledgements}
M.K. acknowledges that the research was supported by the Climate Safety \& Security center of the Delft University of Technology. C.C. acknowledges that the research was supported by the ERC-ADG - Advanced Grant (Grant agreement ID: 833168).

\footnotesize
\bibliographystyle{apalike}
\bibliography{main}

\end{document}